\documentclass[twocolumn,showpacs,preprintnumbers,amsmath,amssymb]{revtex4}
\usepackage{epsfig}
\usepackage{amsmath}

\begin{document}

\title{Knitted Complex Networks}

\author{Luciano da Fontoura Costa}
\affiliation{Institute of Physics at S\~ao Carlos, University of
S\~ao Paulo, PO Box 369, S\~ao Carlos, S\~ao Paulo, 13560-970 Brazil}

\date{8rd Nov 2007}

\begin{abstract}
long paths and hubs
longest path is NP-complete
longest path measurements (length, variations) 
random walks
duality path/star
\end{abstract}

\begin{abstract}
To a considerable extent, the continuing importance and popularity of
complex networks as models of real-world structures has been motivated
by scale free degree distributions as well as the respectively implied
hubs.  Being related to sequential connections of edges in networks,
paths represent another important, dual pattern of connectivity (or
motif) in complex networks (e.g., paths are related to important
concepts such as betweeness centrality).  The present work proposes a
new supercategory of complex networks which are organized and/or
constructed in terms of paths.  Two specific network classes are
proposed and characterized: (i) PA networks, obtained by star-path
transforming Barab\'asi-Albert networks; and (ii) PN networks, built
by performing progressive paths involving all nodes without
repetition.  Such new networks are important not only from their
potential to provide theoretical insights, but also as putative models
of real-world structures.  The connectivity structure of these two
models is investigated comparatively to four traditional complex
networks models (Erd\H{o}os-R\'enyi, Barab\'asi-Albert, Watts-Strogatz
and a geographical model).  A series of interesting results are
described, including the corroboration of the distinct nature of the
two proposed models and the importance of considering a comprehensive
set of measurements and multivariated statistical methods for the
characterization of complex networks.
\end{abstract}

\maketitle

\vspace{0.5cm}
\emph{`You can not travel the path until you have become the path itself.' 
(Gautama Siddharta)}

\section{Introduction} 

\begin{figure}
  \vspace{0.3cm}
  \centerline{\includegraphics[width=1.0\linewidth]{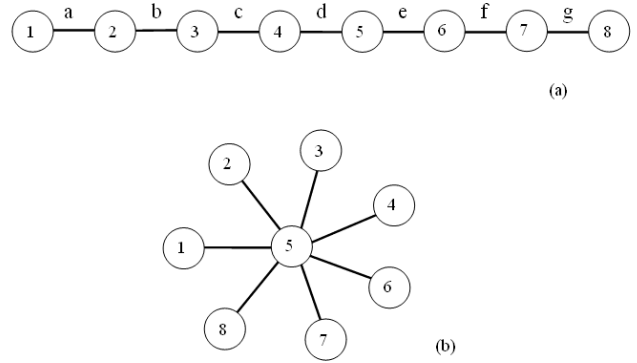}}
  \caption{Examples of simple networks containing a single path (a)
  and a single hub (b).  These two networks have intrinsically
  distinct properties implied by the sequential and centralized nature
  of paths and stars, respectively, which can be considered
  duals.}~\label{fig:path_star}
\end{figure}

Much of the interest in complex networks research
(e.g.~\cite{Albert_Barab:2002, Dorogov_Mendes:2002, Newman:2003,
Boccaletti:2006, Costa_surv:2007}) has been related to particularly
interesting connectivity patterns arising in specific network models.
A small world network, for instance, involves small shortest paths
between its pairs of nodes, and high clustering coefficient.  On the
other hand, a scale free network presents a distribution of node
degrees which follows a power law, enhancing the probability of hubs.
These two types of structures are arguably the most important models
in complex networks research, having been considered by the vast
majority of related works.  Interestingly, while small world networks
are related to shortest paths, scale free structures are intrinsically
linked to the concept of degree and hubs.  It is perhaps not by chance
that the two concepts characterizing these two principal types of
networks have a rather distinct, \emph{dual}, nature.  As a matter of
fact, a path is inerently sequential, while the concept of degree is
associated to the star~\cite{Costa_path:2007} defined by the
connections around a node. Another important issue potentially related
to paths relates to the \emph{causality} and \emph{transitivy} of
effects along the network (especially in the case of oriented
networks).  More specifically, a long oriented path in a network can
be related to a sequence of causal effects.  A star with particularly
high degree has been called a \emph{hub}.  Figure~\ref{fig:path_star}
illustrates two simple networks, one formed by a single path (a) and
another by a single hub.  To a great extent, these two structures
capture the essential features of sequential (e.g. Watts-Strogatz,
which starts as a cycle involving all nodes) and centralized
(e.g. Barab\'asi-Albert) network models.

Although both structures in Figure~\ref{fig:path_star} interconnect
the same set of nodes and contain the same number of edges, they
exhibit a completely distinct, dual, nature.  For instance, the
minimum number of edge crossings implied while visiting all the nodes
can be easily verified to be equal to 7 and 12, respectively (note
that the ratio between these values tends to 2 for large number of
nodes).  This feature corresponds to an inherent advantage of the path
organization.  However, in case the edge d in the structure in
Figure~\ref{fig:path_star}(a) is removed (e.g. through attack or
failure), half of the network (i.e. 1 to 4) will be rendered
unaccessible to the other half (i.e. 5 to 8).  Contrariwise, a removal
of any single edge in the network in Figure~\ref{fig:path_star}(b)
will result in the isolation of just a single node.  This fact
therefore suggests that the centralized network is more resilient to
edge attack, though it is particularly weak to node/hub attack (see,
for instance,~\cite{Albert:2000,Holme:2002,Crucitti:2003}).  The
distribution of shortest path lengths is also markedly different for
each model: ranging between 1 and 7 in the path network and between 1
and 2 for the hub network. The path and star organizations differ in
many other respects, including distribution of betweeness centrality
and node correlations, often leading to opposite properties.

In the light of the above discussion, paths and stars can be
considered dual structural elements (or motifs~\cite{Milo:2002}) in
complex networks.  In particular, long paths would be duals to hubs.
Therefore, it is expected that network models involving many long
paths (e.g. Watts-Strogatz~\footnote{It is important to observe that
though Watts-Strogatz networks are characterized by small average
shortest path length, they do involve at least one long path related
to the initial configuration from which these networks are obtained
(i.e. a cycle involving all nodes).}) will in some way reflect the
basic features of the path motif, i.e. their nodes can be effectively
covered by relatively short path walks~\footnote{A \emph{path-walk} is
a walk which never visits any of the nodes or edges more than once.
Such walks can also be called self-avoiding walks.}, but they will
present low resilience to edge attack.  Contrariwise, networks
organized around hubs (e.g. Bara\'asi-Albert structures) would be
expected to be resilient to edge attack, but imply long path walks
required to cover most of their nodes.

The identification of the duality between the path/star motif as well
as between networks involving long paths and large hubs paves the way
to several investigations in complex network research.  These
possibilities include but are not limited to: (i) comprehensive
investigations of the topological properties of networks involving
paths or hubs; and (ii) the definition of transformations between such
motifs and networks and study of the respective effects on the
networks properties.  The latter possibility was pre;iminarly explored
in a recent work~\cite{Costa_path:2007}, where the path-star and
star-path transformations were proposed.  That work considered the
transformation of just one path, starting from a randomly chosen
node. The effects of such transformation on the overall network
connectivity were quantified in terms of ratios between measurements
such as the average clustering coefficient and average shortest path
length after and before the transformations.  While the path-star
transformation tended to increase the average clustering coefficient
in Erd\H{o}s-R\'enyi (ER) networks, it implied small changes in
Barab\'asi-Albert (BA) structures.  Major decreases of the average
shortest path length were observed for Watts-Strogatz (WS) networks.

\begin{figure}
  \vspace{0.3cm}
  \begin{center}
    \includegraphics[width=0.9\linewidth]{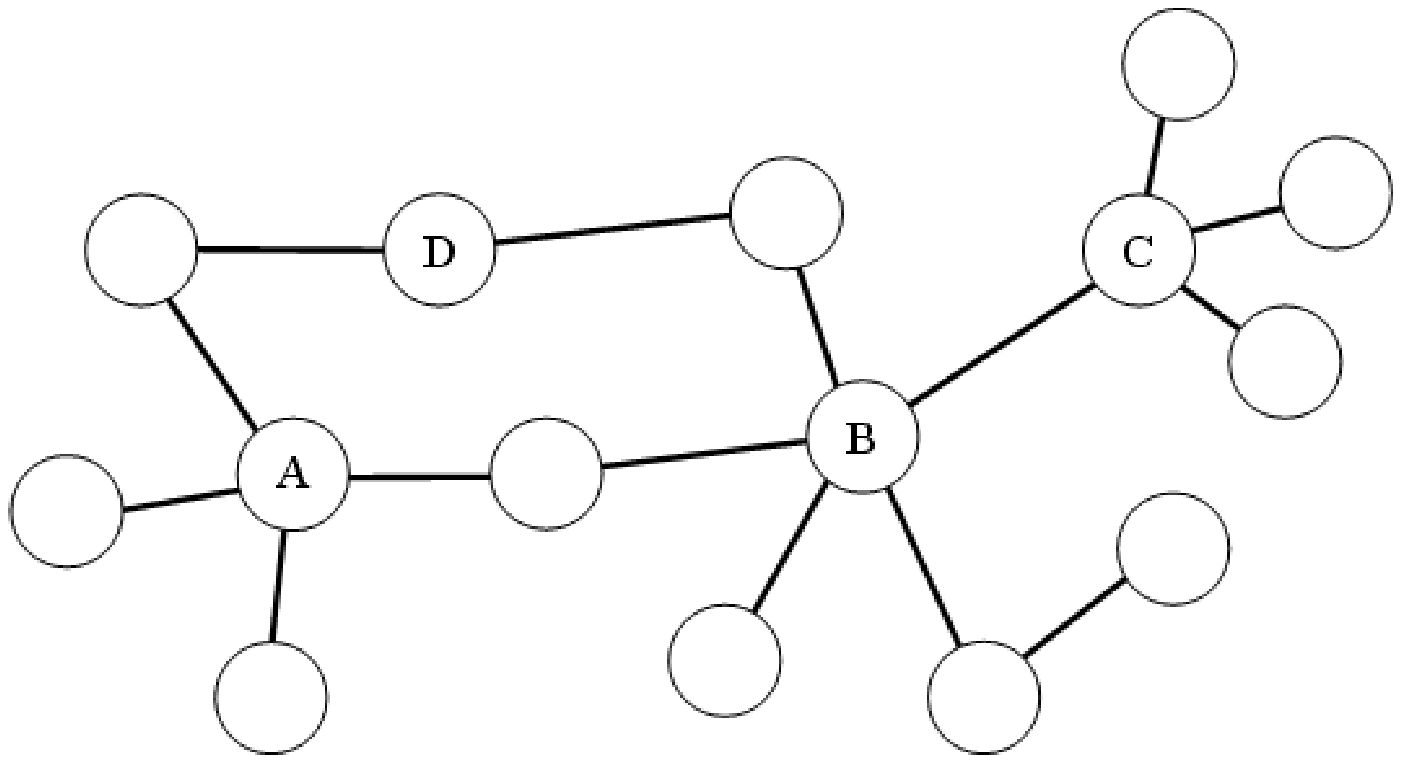} \\
    (a) \\ 
    \includegraphics[width=0.9\linewidth]{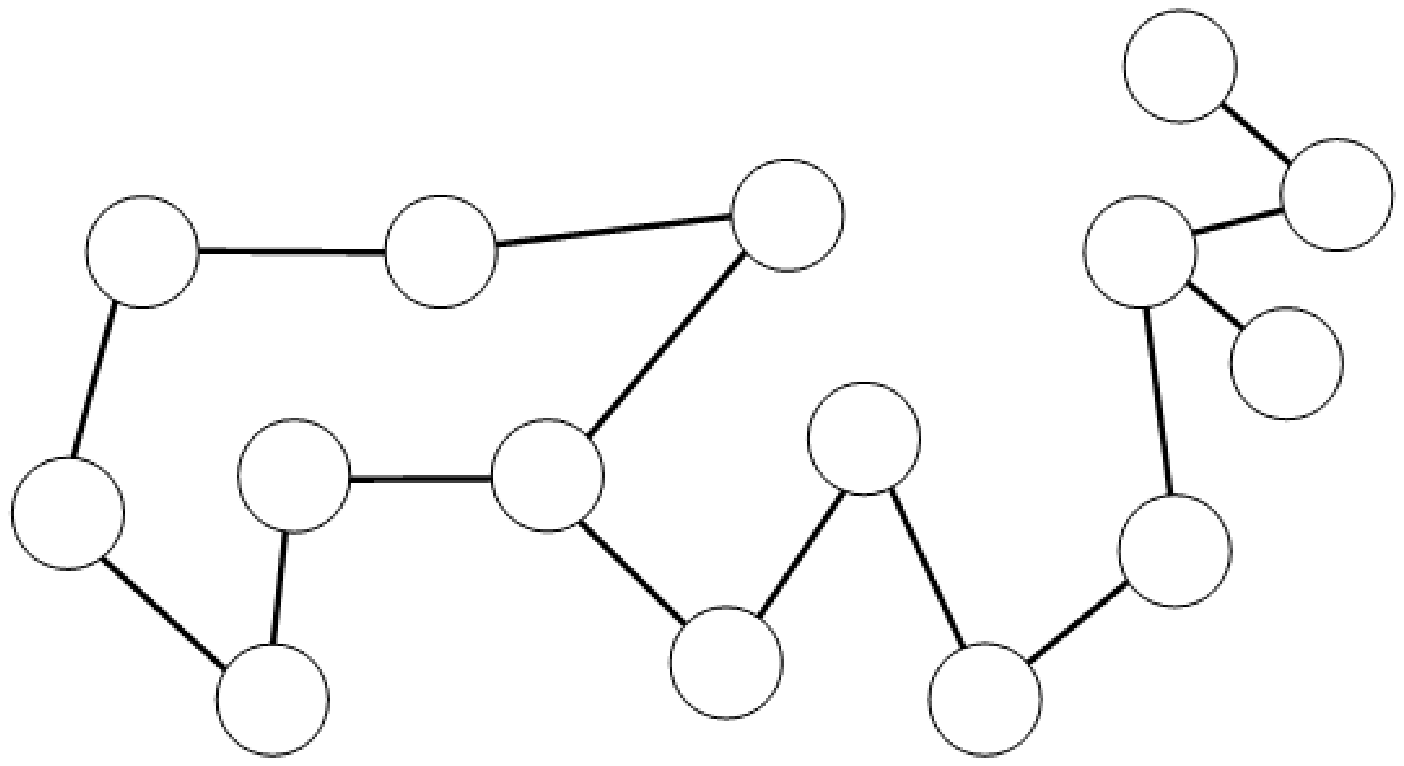} \\
    (b) \\ 
    \includegraphics[width=0.9\linewidth]{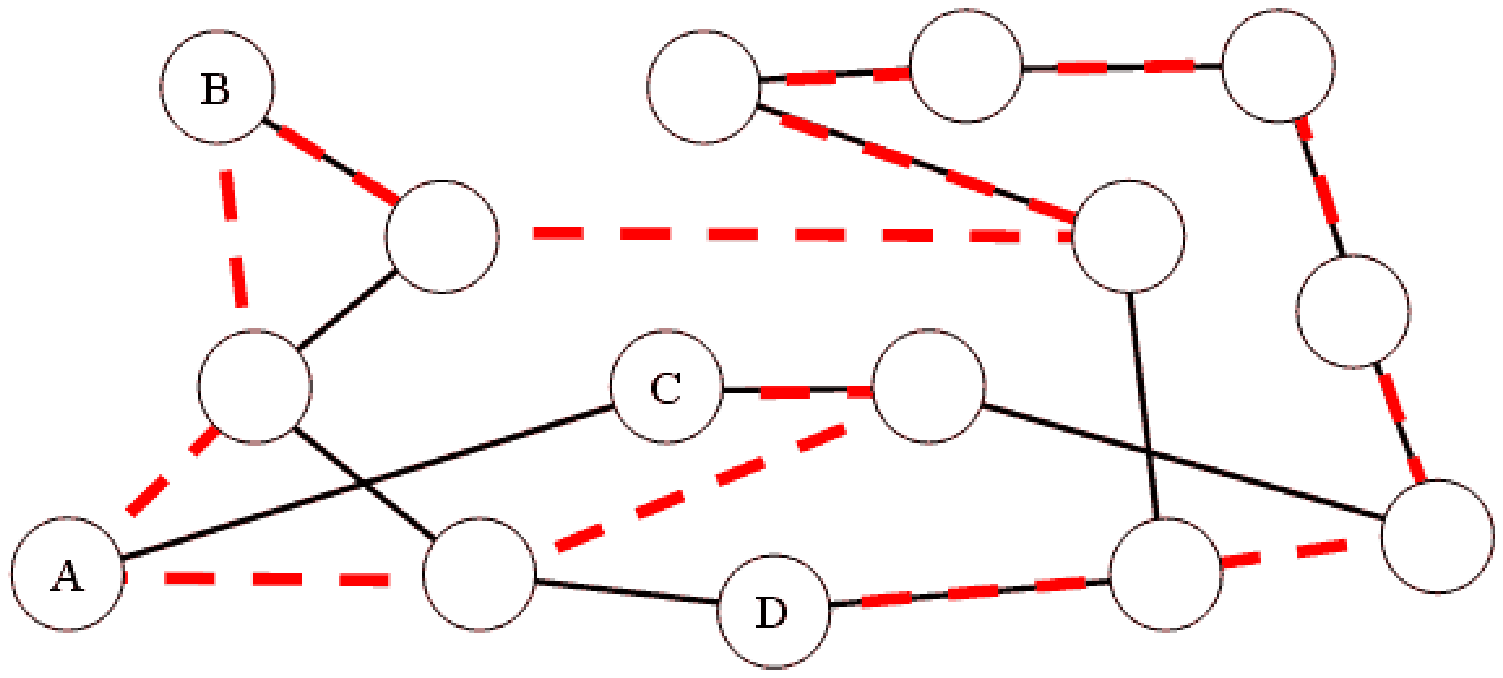} \\
    (c)
    \caption{Examples of knitted networks.  A simple network involving
    hubs is shown in (a), while one of its possible path-transformation 
    is shown in (b).  Both these networks have exactly the same number of 
    nodes and edges, but exhibit completely distinct topologies.  A small PN
    network is shown in (c). This network was obtained by
    incorporating two path-walks, one initiating in A and terminating
    in B and another (in red) initiating in C and ending in D.  Observe that
    most nodes have degree equal to 4 and are visited exactly once by
    each of the two path-walks.  Observe that it is straightforward
    to obtain directed version of PN networks, with the directions
    defined by the sense of the path-walk.}~\label{fig:exs}
  \end{center}
\end{figure}

The present work continues and extends the previous investigation
in~\cite{Costa_path:2007} by addressing the characterization of two
networks grown in terms of paths, which are henceforth called
\emph{knitted networks} because of the similarity of their growth with
the progressive incorporation of treads (i.e. paths).  These two novel
types of complex networks include: (a) networks obtained by the
star-path transformations of all stars in BA networks; and (b)
networks whose edges are obtained by randomly selecting each of the
nodes just once.  These two categories of networks, illustrated in
Figure~\ref{fig:exs}, are henceforth called \emph{Path-transformed BA}
(PA) and \emph{Path-regular} (PN) networks.  ER, BA, WS and a
geographical type of networks have also been considered for
comparative purposes.

Regarding the path-transformed BA (PA) model, it is obtained by
transforming each of the stars in a BA network into a respective path.
The transformation from star to path is performed in decreasing order
of node degree, so that hubs are transformed first. The edges removed
from the stars in the original BA network are progressively
transferred to the growing transformed network.  Thus, each of the
paths incorporated into the PA network corresponds to a respective
star in the original BA structure.  As such, the PA network has an
embedded distribution of paths which exhibits the same power law as
the degrees of the original BA network.  For such a reason, this new
kind of network can be thought as the path-dual of the BA model.  

It is important to observe that the star-path transformation is not
deterministic, in the sense that several slightly distinct networks
can be obtained by transforming a star into a path (as explained in
more detail in Section~\ref{sec:PA}, this is a consequence of the
fact that there is no clear way in which to choose the sequence of
nodes from the hub while generating the path, so that they are
randomly selected.).  On the other hand, the path-star transformation
is deterministic.  

The other considered knitted model, namely the path-regular (PN)
structures, represents a simpler but nevertheless potentially
interesting new model as it exhibits an almost perfect degree
regularity (in the sense of most nodes exhibiting the same degree).
As presented in this article, this network also resulted remarkably
regular with respect to several other measurements.  A PN network is
obtained by connecting all the nodes (initially isolated) so that
every node is selected only once, therefore defining a path in the
growing network.  This procedure can be repeated several times. The
degrees of most nodes therefore result very close to twice the number
of complete paths involving the whole set of nodes.

The two main reasons why it is interesting to consider new network
models such as those outlined above are: (i) such models can provide
theoretical insights about aspects related to their intrinsic nature
or to the way in which they are grown; and (ii) it is useful to
compare such models with specific real-world networks, in the sense
that good similarity between them may contribute to understanding
real-world problems.  The proposal of novel types of networks thus
immediately implies the important question of how such structures
relate to other existing models, especially those more traditional
such as uniformly random, scale-free, small world and geographical.
It is therefore important to devise a reasonable methodology allowing
such a problem to be properly tackled. The experimental methodology
adopted in this article involves generating several realizations of
each category of network, for two sizes (i.e. $N=100$ and $N=200$) and
two average degrees (i.e. $\left< k \right> = 6$ and $\left< k \right>
= 10$), and calculating a series of distinct measurements of their
respective topology.  The six networks types are them compared in
terms of projections of these measurements, obtained by using
canonical analysis, which projects the measurement space so as to
maximize the separation of the six network categories
(see~\cite{Costa_surv:2007}).

This article starts by presenting the basic concepts and methods in
complex network, their growth, measurements, and multivariate methods
for projection and analysis of the similarities between the models.
It follows by presenting the obtained results and respective
discussion.

\section{Basic Concepts and Methodology}

This section describes the basic concepts and methods used in the
present work, including the representation and measurement of
networks, the traditional complex network models (ER, BA, WS and a
geographical type of network), and multivariated statistical methods,
which are described in the respective subsections.  Additional
information about network measurements and their analysis by using
statistical methods can be found in~\cite{Costa_surv:2007}.

\subsection{Network Representation and Measurements}

Complex networks are discrete structures involving $N$ nodes and $E$
edges connectiong those nodes.  In this work we focus attention on
undirected networks.  This type of network can be represented by a
symmetric adjacency matrix $K$ such that the presence of each edge
$(i,j)$, where $i$ and $j$ are any network nodes and $i\neq j$,
implies $K(i,j)=K(j,i)=1$, with $K(i,j)=K(j,i)=0$ otherwise. 

A \emph{star} in a network is henceforth understood as any node
together with its respectively attached edges.  Two edges $(i,j)$ and
$(k,m)$ are ajacent whenever they share an extremity (i.e. $i=k$ or
$j=k$ or $i=m$ or $j=m$).  A sequence of adjacent edges constitues a
\emph{walk}.  Observe that a walk may go more than once over the same
nodes or edges. A \emph{path} is a walk which never re-visit any edge
or node.  A closed path is \emph{cycle}.  The \emph{diameter} of a
network correspond to the length of the longest shortest path between
any pair of nodes.

The \emph{immediate neighbors} of a node $i$ are those which are
distant by one edge from $i$.  The degree of a node is equal to the
number of its immediate neighbors.  The
\emph{clustering coefficient} of a node $i$ quantifies how well the 
immediate neighbors of that node are interrelated.  More specifically,
if $n(i)$ is the number of immediate neighbors of node $i$, then its
clustering coefficient can be calculated as:

\begin{equation}
  cc(i) = \frac{2e(i)}{n(i)(n(i)-1)}
\end{equation}

where $e(i)$ is the total number of undirected edges connecting the
immediate neighbors of $i$.  Though the degree and the clustering
coefficient, which are traditionally adopted measurements, are defined
for each node in a network, it is also interesting to consider the
\emph{average} and \emph{standard deviation} of those values as global 
features of the whole network.  

The clustering coefficient can be generalized to consider sets of
nodes other than the immediate neighbors of a reference
node~\cite{Costa_EPJB:2005,Costa:2004}. In this work we use the second
order clustering coefficient --- \emph{second clustering coefficient},
for short --- of each node $i$, $cc2(i)$, which considers the
interconnectivity of the second neighbors of that node.  The second
neighbors are those nodes which are can be accessed from node $i$ by a
shortest path of length 2.  The second clustering coefficient can be
calculated as:

\begin{equation}
  cc2(i) = \frac{2e2(i)}{n2(i)(n(i)-1)}
\end{equation}

where $e2$ is the number of undirected edges among the second
neighbors of node $i$ and $n2(i)$ is the number of those neighbors.
Similarly to the degree and clustering coefficient, it is also
interesting to consider the average and standard deviation of the
second clustering coefficient.

Table~\ref{tab:meas} presents all the 9 measurements considered in the
present work and their respective abbreviations.

\begin{table}[htb]
\centering
\vspace{1cm}
\begin{tabular}{|c||c|}
  \hline  
          Measurement                     &   Abbreviation  \\ \hline \hline
          Average Node degree             &   $\left< k \right>$ \\ \hline     
          St. deviation of node degree    &   $\sigma_k$         \\ \hline
          Average clustering coefficient  &   $\left< cc \right>$ \\ \hline
          St. deviation of clust. coeff.  &   $\sigma_{cc}$         \\ \hline
          Average second clust. coeff.    &   $\left< cc2 \right>$ \\ \hline
          St. deviation clust. coeff      &   $\sigma_{cc2}$       \\ \hline
          Network diameter                &   $D$                  \\ \hline
          Average shortest path length    &   $\left< sp \right>$  \\ \hline
          St. deviation short. path leng. &   $\sigma_{sp}$  \\ \hline
  \hline
\end{tabular}
\caption{The measurements of the network connectivity considered 
in the present work and their respective abbreviations.}\label{tab:meas}
\end{table}

\subsection{Traditional Complex Networks Models}

In addition to the two theoretical models of networks proposed in this
article (see Section~\ref{sec:newmodels}), four traditional models are
also considered in order to provide comparison references.  The basic
procedure to obtain these three models --- namely Erd\H{o}s-R\'enyi
(ER), Barab\'asi-Albert (BA), Watts-Strogatz (WS) and a simple
geographical model (GG) --- are described as follows (see
also~\cite{Albert_Barab:2002, Dorogov_Mendes:2002, Costa_surv:2007,
Boccaletti:2006, Newman:2003}.

ER networks are obtained by starting with $N$ nodes and establishing
undirected connections with fixed probability $\gamma$.  The BA
structures are grown from $m0$ initial, randomly connected nodes,
through the progressive addition of new nodes with $m$ edges each,
which are connected with the previous nodes preferentially to their
node degrees.  The WS networks are obtained by starting with a cycle
containing all the $N$ nodes and then rewiring a proportion $\alpha$
of the undirected edges.  The geographical model adopted in this work,
henceforth abbreviated as GG, involves distributing the $N$ nodes with
uniform random probability within a square two-dimentional space and
then connecting each node (through undirected edges) to all nodes
which are no longer than a maximum Euclidean distance $d_{max}$.  All
considered networks are undirected and do not include
self-connections.  The also have the same number of nodes $N$ and
similar average degrees $\left< k \right>$, which are determined in
terms of $m$ (see, for instance,~\cite{Costa_trails:2007,
Paulino:2007}).

\subsection{Multivariated Statistical Methods}

The set of $M$ measurements obtained for a complex networks under
analysis can be represented in terms of a \emph{feature vector} in
$R^M$.  The so-defined $M-$dimensional space is frequently called the
\emph{measurement space} adopted in a specific investigation.  Therefore, 
each distinct network is mapped into a point, defined by the
respective feature vector, in the measurement space.  Observe that
this mapping is not invertible because more than one network, though
structurally distinct, may be mapped into identical feature vectors.
Groups of points (also called clusters) appearing in such spaces
indicate possible categories of networks exhibiting similar
topological properties.  Networks which are mapped into relatively
distant points can be understood to present substantial differences in
their topology (e.g.~\cite{Costa_surv:2007}).

Because the number of adopted measurements $M$ is frequently larger
than 2 or 3, it becomes impossible to visualize the distribution of
the networks when mapped into the $M-$dimensional measurement space.
Fortunately, it is possible to use stochastic projections in order to
reduce the dimensionality of such spaces.  In the present work we
consider the \emph{canonical projection}
(e.g.~\cite{Costa_surv:2007}), where the projection is performed so as
to maximize the distance between the categories of networks while
simultaneously minimizing the dispersions inside each category.

\section{Knitted Network Models}~\label{sec:newmodels}

The two novel categories of networks proposed and investigated in this
work are the path-transformed BA (PA) and the path-regular (PN)
models.  These are here understood to belong to a new supercategory of
structures called \emph{knitted networks}, characterized by being
formed by paths.  These two models are described in the next two
subsections, respectively.

\subsection{Path-Transformed BA Networks (PA)}~\label{sec:PA}

Given a generic complex networks, it is possible to transform each of
its stars into a path by using a methodology recently
suggested~\cite{Costa_path:2007}.  Basically, given a node $i$ and its
connected edges (i.e. a star) in the original network, the following
steps are performed: (i) the set of nodes $S$ comprising the node $i$
as well as its immediate neighbors is identified; (ii) one of these
nodes, except $i$, is randomly chosen as the beginning of the path;
(iii) the edges interconnecting $i$ to its immediate neighbors are
randomly selected and removed from the original network and used to
define the continuation of the growing path in the new network.
Oberve that the edges are transported from the original network into
the growing network, i.e. two networks are considered in order to
avoid transporting one edge more than once.  In order to preserve the
total number of edges, the transported edges are added to eventual
edges already transported.  Therefore, the transformed network is, in
principle, a weighted structure (however, the present work considers
all weights equal to one).

Because each star-path transformation depends on the choice of the
sequence of nodes to be used along the path, it is important to
investigate the intensity of the effects that such variations may have
on the overall network structure. The average and standard deviation
values of the clustering coefficients obtained for 100
path-transformations of the same initial BA network are: $0.045 \pm
0.011$ for $N=100$ and $m=3$ (coefficient of variation = (standard
deviation)/(average) = 0.24), $0.0220 \pm 0.0048$ for $N=200$
(coefficient of variation 0.22) and $m=3$, $0.0497 \pm 0.0041$ for
$N=100$ and $m=5$ (coefficient of variation 0.08), and $0.045 \pm
0.0048$ for $N=200$ and $m=5$ (coefficient of variation 0.11).  These
results suggest that, at least for the considered configurations and
especially for larger values of $m$, the structural changes implied by
the diverse choices of the nodes for the star-path transformation tend
to lead to relatively small effects on the connectivity properties of
the diverse PA networks obtained from the same BA structure.

\subsection{Path-Regular BA Networks (PN)}

These networks are simply constructed by starting with $N$ isolated
nodes and incorporating $M$ sequences of edges defined by choosing
non-repeated, randomly chosen nodes (all nodes are taken).  See
Figure~\ref{fig:exs}(c) for an illustration of a PN network built up
by choosing the $N$ nodes according to two random sequences.  Thus,
this type of networks is obtained, as hinted in the quotation in the
beginning fo this text, the networks are obtained from paths, not
vice-versa, i.e. the paths and the network become one.  Because each
path enters and leaves most of the nodes $2$ times, the average node
degree in this category of networks tends to converges steadily to
$\left< k \right> = 2M$.  Therefore, this network is highly regular at
least with respect to the node degree.  After random networks such as
those in the ER model, the PN networks are possibly those which are
simplest to grow computationally.

\section{Results and Discussion}

A total of 50 simulations were performed for the PA, PN, ER, BA, and
WS models with respect to each of the four following configurations:
(a) $N=100$ and $m=3$; (b) $N=100$ and $m=5$; (c) $N=200$ and $m=3$
and (d) $N=200$ and $m=5$.  The obtained results are presented and
discussed in the following sections with respect to several
progressive combinations of measurements and projections. 

Only the largest connected component of the original networkz are
considered (note that, because of the relatively high average degrees
considered in this work, the largest component often corresponds to
the whole original network).

\subsection{Node Degree and Clustering Coefficient}\label{sec:deg_cc}

\begin{figure*}
  \vspace{0.3cm}
  \begin{center}
  \includegraphics[width=0.45\linewidth]{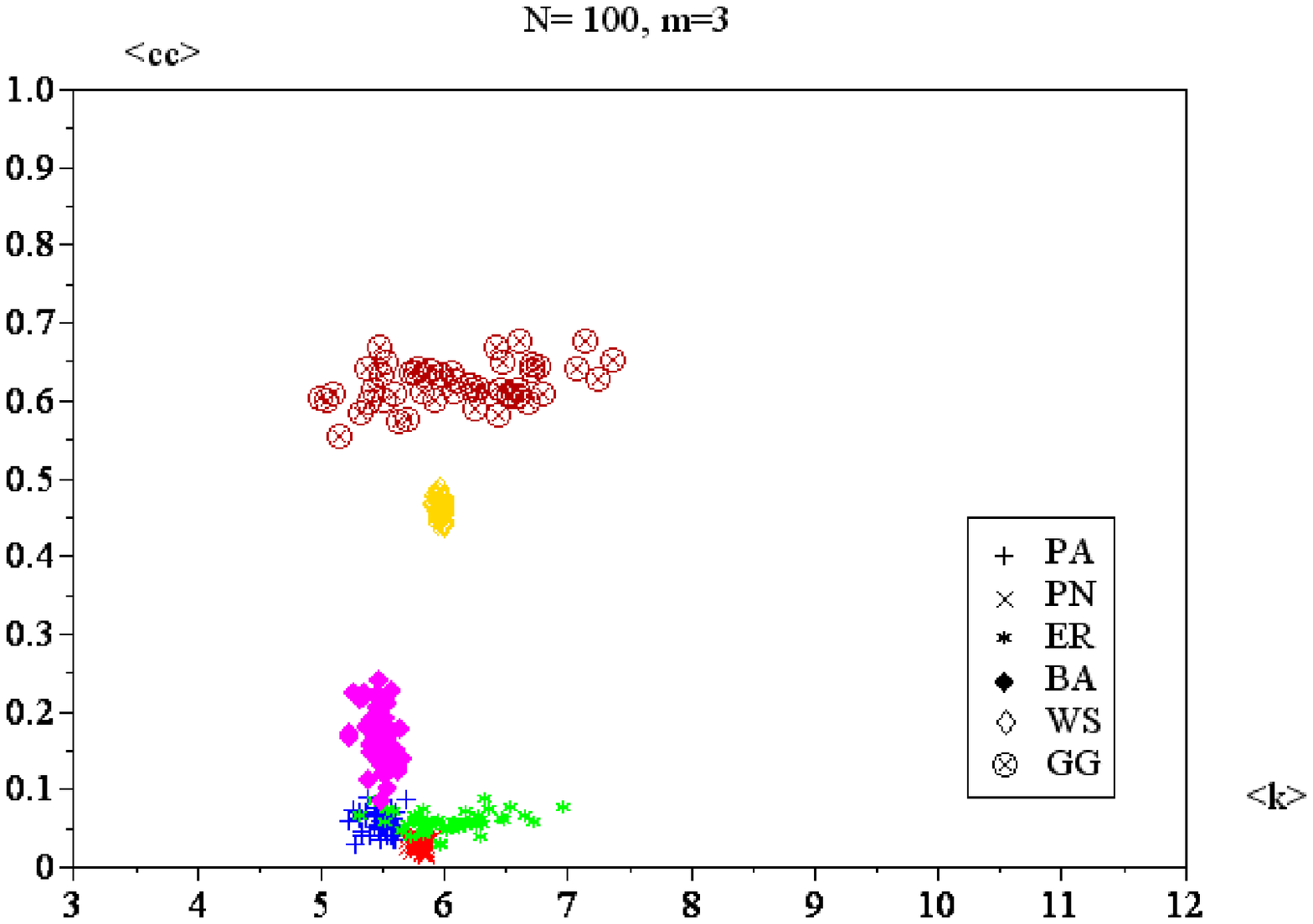}  \hspace{0.5cm}
  \includegraphics[width=0.45\linewidth]{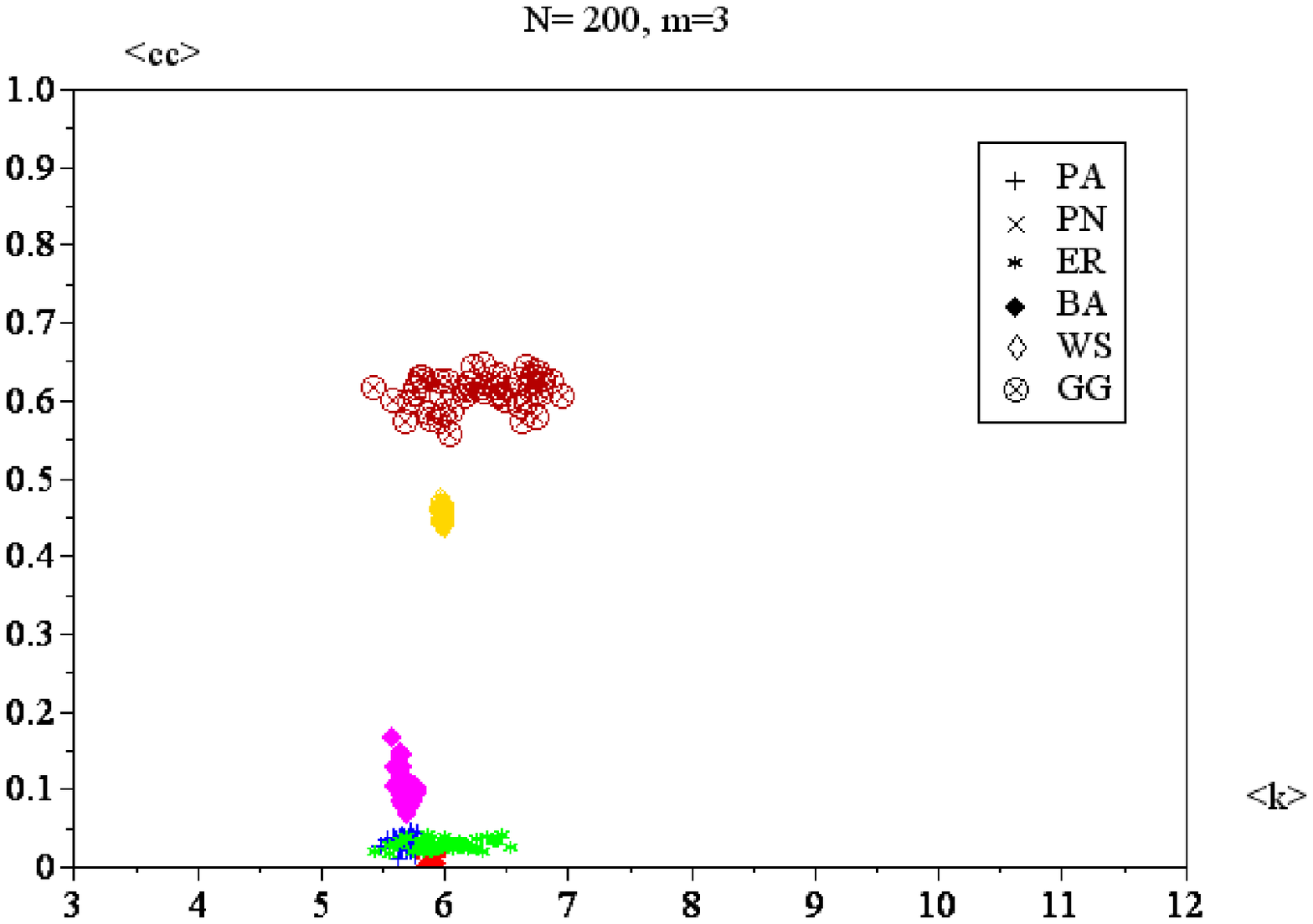}  \\
   (a)  \hspace{7cm}  (b) \\
  \includegraphics[width=0.45\linewidth]{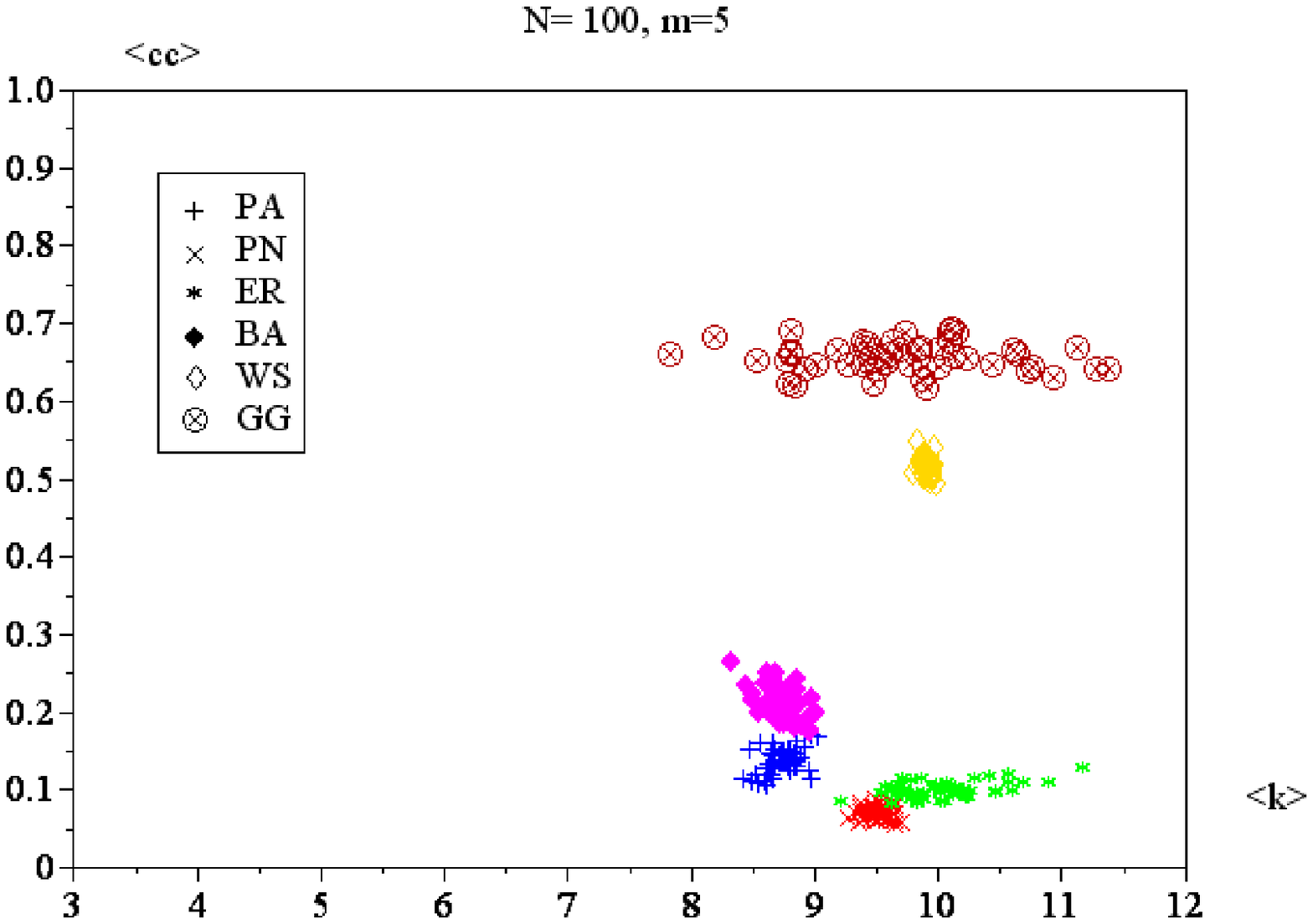}  \hspace{0.5cm}
  \includegraphics[width=0.45\linewidth]{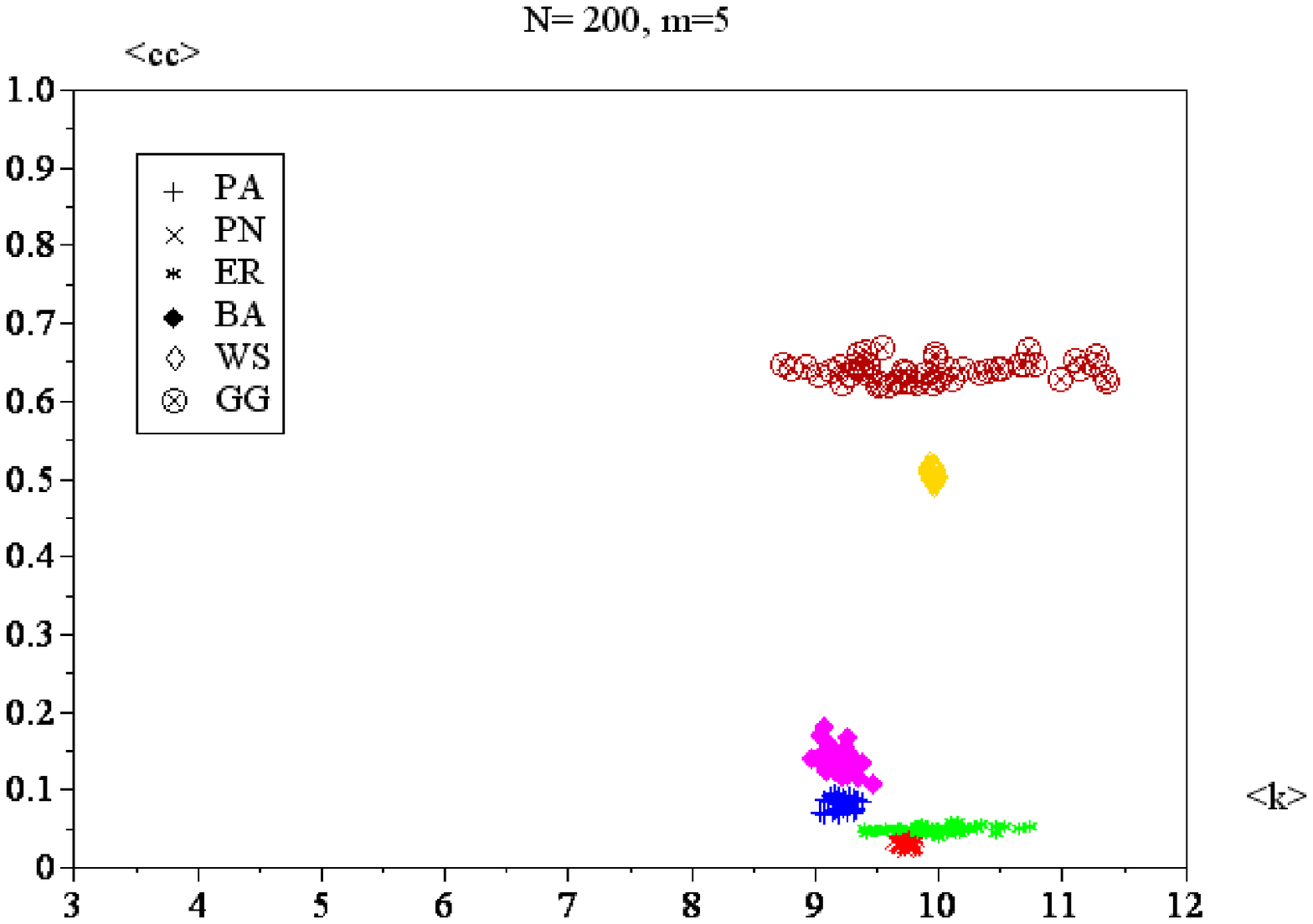}  \\
   (c)  \hspace{7cm}  (d) \\
  \caption{The measurement spaces involving the average node degree
  and average clustering coefficient obtained for the 50 instances of
  each of the 5 network models considered in this article with respect
  to the four configurations: (a) $N=100$ and $m=3$; (b) $N=100$ and
  $m=5$;(c) $N=200$ and $m=3$; and (d) $N=200$ and $m=5$.  See text
  for discussion.}~\label{fig:deg_cc}
  \end{center}
\end{figure*}

The average values of the node degree and clustering coefficient have
been traditionally used in order to quantify the topological
properties of complex networks.  We start our investigation of the
structural relationship between the new PA and PN networks with the
traditional ER, BA, WS and GG models by considering these two
measurements.  The distributions of the several instances of each
network category along this two-dimensional measurement space,
considering the four adopted configurations, are shown in
Figure~\ref{fig:deg_cc}.

A series of interesting facts can be inferred from these figures.
First, note that, as expected, all nodes have average degrees near
$2m$ (i.e. $\left< k \right> \approx 6$ for $m=3$ and $\left< k
\right> \approx 10$ for $m=5$).  Observe that, because of finite size 
effects, the average degrees of the BA networks are smaller than the
expected theoretical value for $m=3$, but tends to approach $2m$ for
the larger value $m=5$. A similar tendency can be observed for the PA
model, which is derived from the BA counterparts.  Rather distinct
dispersions of the average node degree are observed for each model,
with the WS resulting in the smallest variation, while the GG model
accounts for the largest dispersion.  Interestingly, the PA model
exhibited degree dispersion very similar to that of the BA model.  The
average values of clustering coefficients presented much less
dispersion than the average degree.  As expected, each type of network
resulted with typical and characteristic average clustering
coefficient values.  The smallest clustering coefficients were
obtained for the ER, PA and PN models, while the WS and GG models
presented the highest values.  Intermediate values of clustering
coefficient were obtained for the BA networks.  The dispersions of
both the average degree and average clustering coefficient tend to
decrease with $N$.

To judge from the measurement spaces defined by the average degree and
average clustering coefficient, the PN novel type of network tends to
be similar to the ER model (however, the ER networks present
substantially higher node degree variance).  At the same time, the PA
model presented average degree similar to the BA, but smaller average
clustering coefficients (the average clustering coefficient of the PA
networks is comparable to those of the ER structures).  Generally, if
only these two measurements are considered, we can conclude that the
PN and PA novel networks are reasonably similar to the ER and BA types
of networks, respectively.  As a matter of fact, because such
differences are mostly related to the average degree (which tends to
become more similar for large values of $N$), \emph{these four
networks could ultimately be understood (incorrectly, as it will
become clear soon) to exhibit little structural differences.}.

\begin{figure*}
  \vspace{0.3cm}
  \begin{center}
  \includegraphics[width=0.45\linewidth]{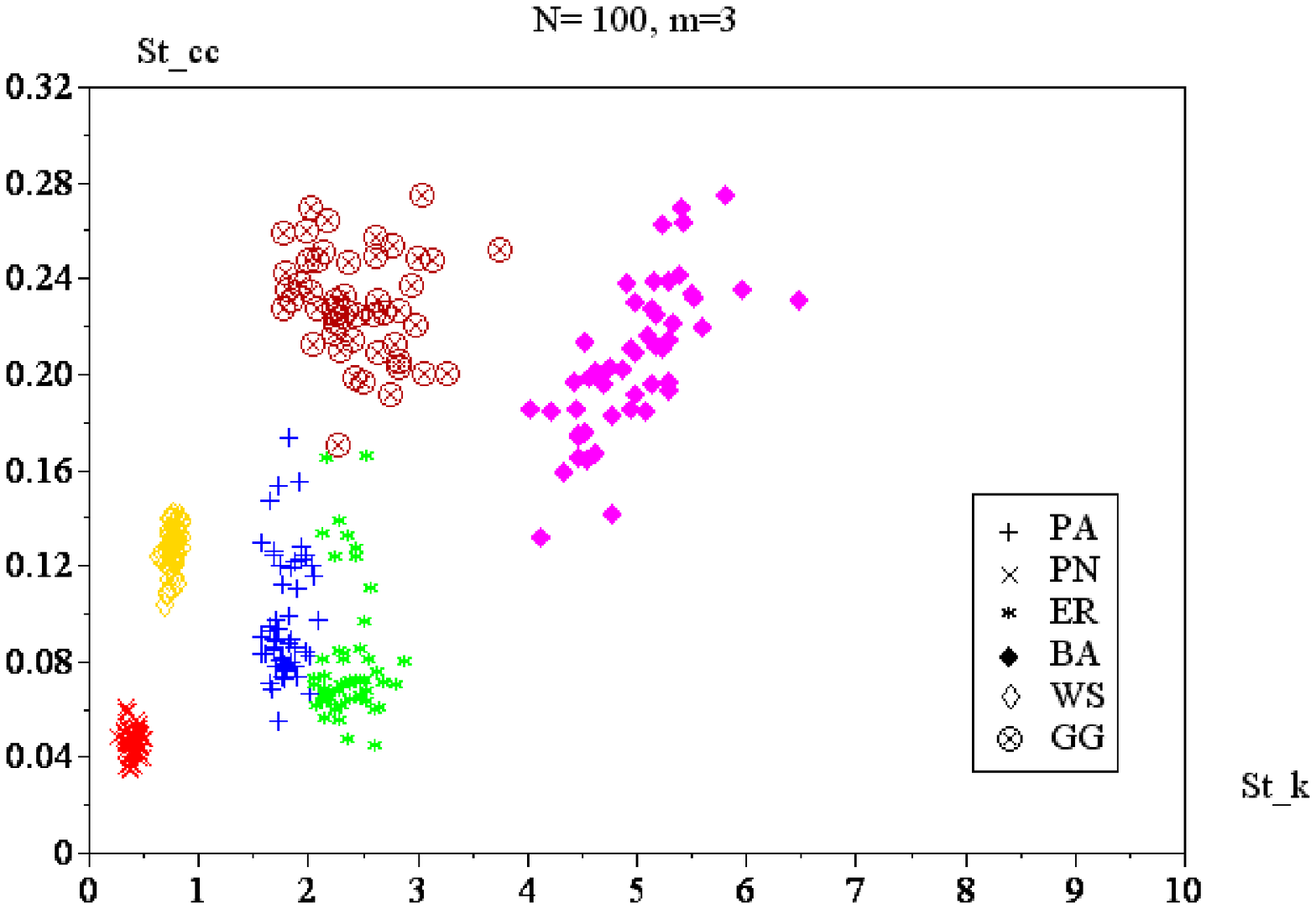}  \hspace{0.5cm}
  \includegraphics[width=0.45\linewidth]{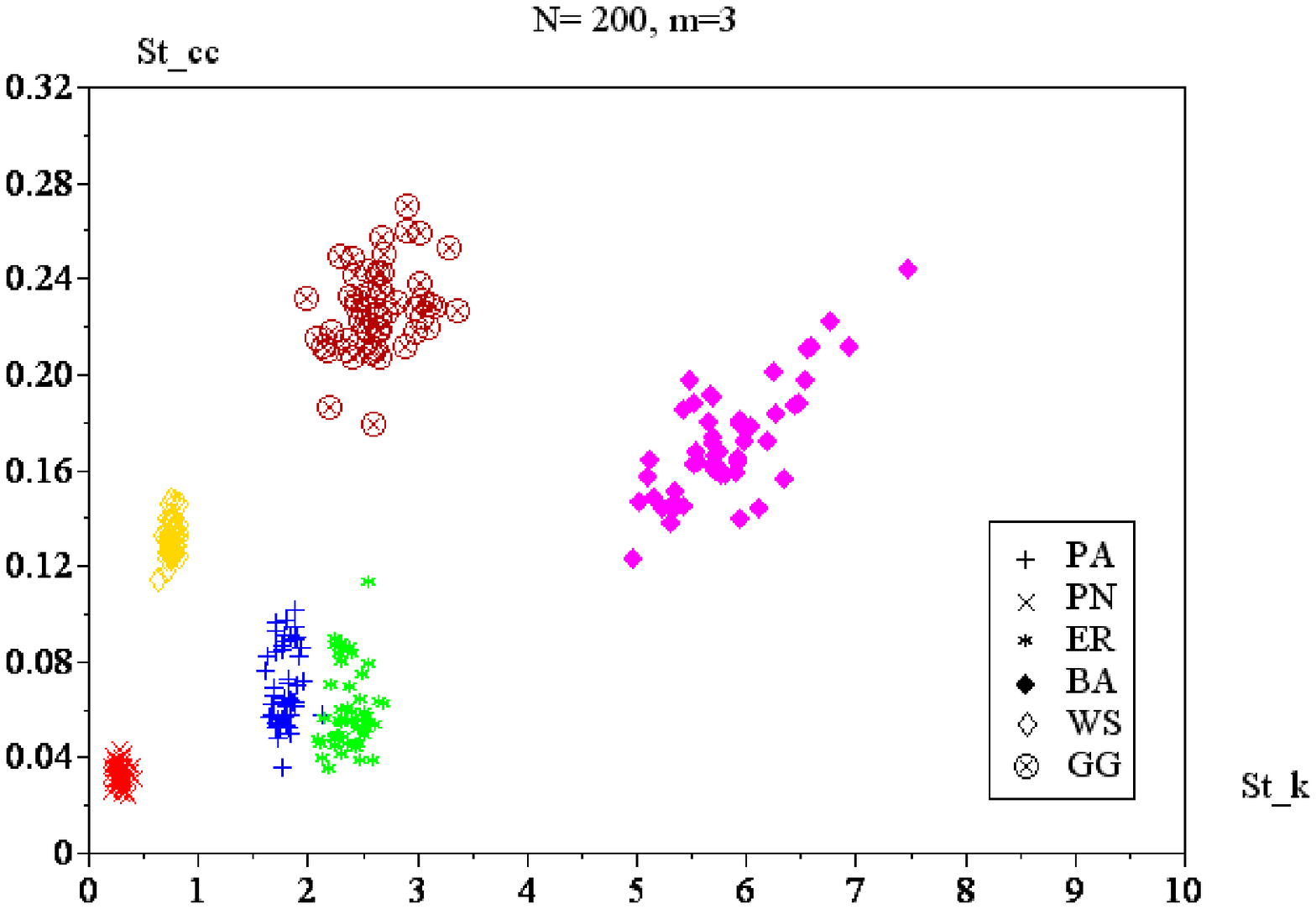}  \\
   (a)  \hspace{7cm}  (b) \\
  \includegraphics[width=0.45\linewidth]{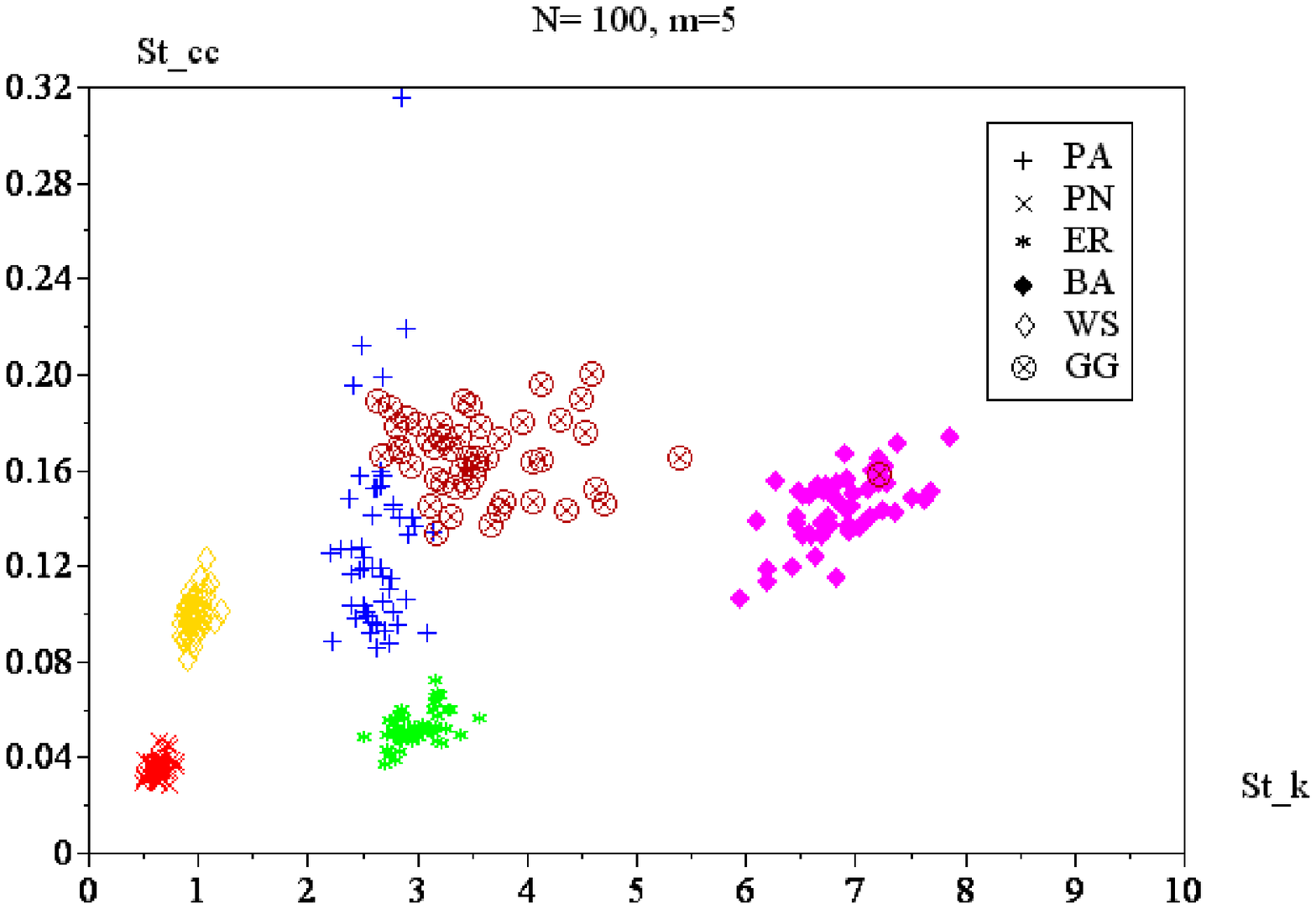}  \hspace{0.5cm}
  \includegraphics[width=0.45\linewidth]{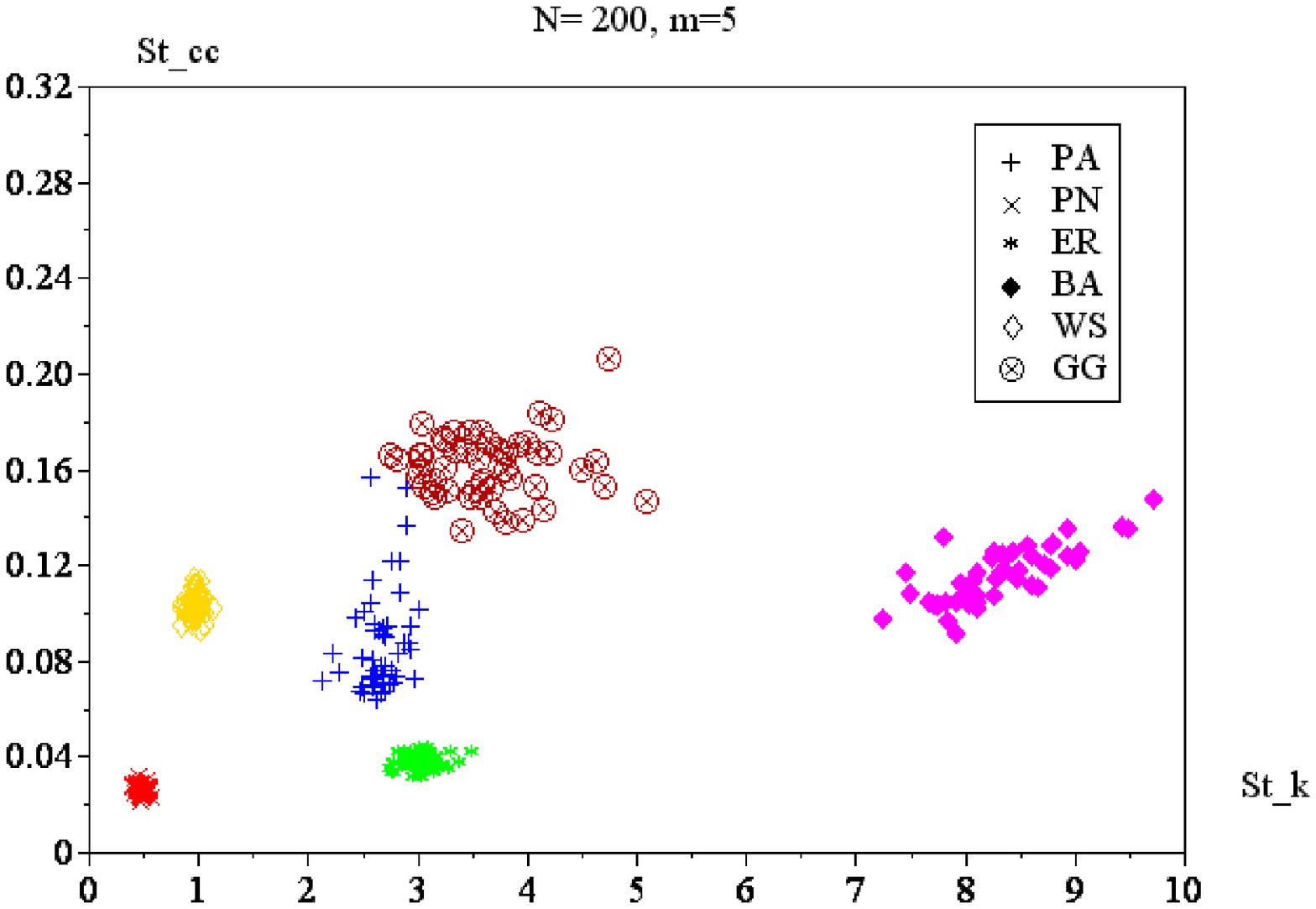}  \\
   (c)  \hspace{7cm}  (d) \\
  \caption{The measurement spaces involving the standard deviations of
  the node degree and clustering coefficient obtained for the
  50 instances of each of the 5 network models considered in this
  article with respect to the four configurations: (a) $N=100$ and
  $m=3$; (b) $N=100$ and $m=5$;(c) $N=200$ and $m=3$; and (d) $N=200$
  and $m=5$.}~\label{fig:st_deg_cc}
  \end{center}
\end{figure*}

Figure~\ref{fig:st_deg_cc} shows the distribution of the networks
mapped in measurements spaces defined by the standard deviations of
the respective node degrees and clustering coefficients.  Observe that
such standard deviations are calculated considering the degrees and
clustering coefficients of each individual node in each of the network
realizations.  Though naturally implied, such measurements have been
rarely used in the complex network area.

It is clear from these scatterplots that a substantially more distinct
separation between the 6 complex network types is now obtained.  A
series of interesting trends can be discerned from
Figure~\ref{fig:st_deg_cc}.  First, as before, the overall dispersion
of the clusters tended to decrease with $N$, for a fixed $m$.  Also,
the BA networks presented, in all cases, the largest standard
deviations of node degrees (a fact implied by the wider distribution
of node degrees in that model, with presence of hubs), while the GG
model yielded the largest standard deviations of clustering
coefficients in all situations (an interesting effect related to the
distribution of nearest distances in systems of uniformly distributed
points).  As expected, the WS structures provided reasonably small
dispersions of both degree and clustering coefficient, therefore
defining a well-separated cluster in all scatterplots in
Figure~\ref{fig:st_deg_cc}.  Recall that we are referring to the
coordinate values expressing the standard deviations, not the
intrinsic dispersion of the clusters in the scatterplots. 

Unlike in the previous analysis involving the average degree and
average clustering coefficient, the novel PN model resulted
well-separated from all other network categories, presenting the
overall smallest standard deviations of both degree and clustering
coefficient values in all scatterplots.  This fact suggests an
\emph{intense regularity} of the PN model (please refer 
to~\cite{Costa_seeking:2007} for a discussion on the generalization of
the concept of regularity to topological measurements other than the
node degree), in the sense that most nodes in that type of network
present similar degree and clustering coefficient values.  Unlike in
our previous analysis involving the average values of degree and
clustering coefficient, the new PA model now resulted well away from
the BA model and more distinct from the ER model (except for the case
$N=100$ and $m=3$ in Figure~\ref{fig:st_deg_cc}).  Such results
clearly indicate that the PN and PA models do exhibit specific
structural features which distinguish them from the other
traditionally used measurements.  In particular, the PN stands out as
being particularly regular with respect to both node degree and
clustering coefficient.  On the other hand, the PA networks tend to be
similar to the ER model, particularly for smaller values of $N$ and
$m$.  However these two types of networks tend to separate one another
for larger values of those parameters.  For the values of $N$ and $m$
considered in this work, it is reasonable to say that the ER is the
model which is closest to the PA as far as the standard deviations of
the degree and clustering coefficient are concerned.

\subsection{All Together Now}

In order to gain further insights about the possible structural
relationships between the novel PN and PA networks and the four
traditional models considered in this work, we now resource to all the
masurements in Table~\ref{tab:meas}.  Observe that a single value of
diameter is obtained for the whole network, so that no average or
standard deviation values are considered for this global feature.
Because of the high dimensions of the the measurement spaces implied
by these several measurements, it is no longer possible to visualize
the distribution of the networks in the respectie measurement spaces,
as done in the previous section.  In order to circumvent this problem,
in the following we apply the canonical projection methodology to
those measurement spaces.  In addition, because the PN and PA models
have already been found to be considerably distinct from the WS and GG
models, these two traditional theoretical models are no longer
considered in the following analysis.

\begin{figure*}
  \vspace{0.3cm}
  \begin{center}
  \includegraphics[width=0.4\linewidth]{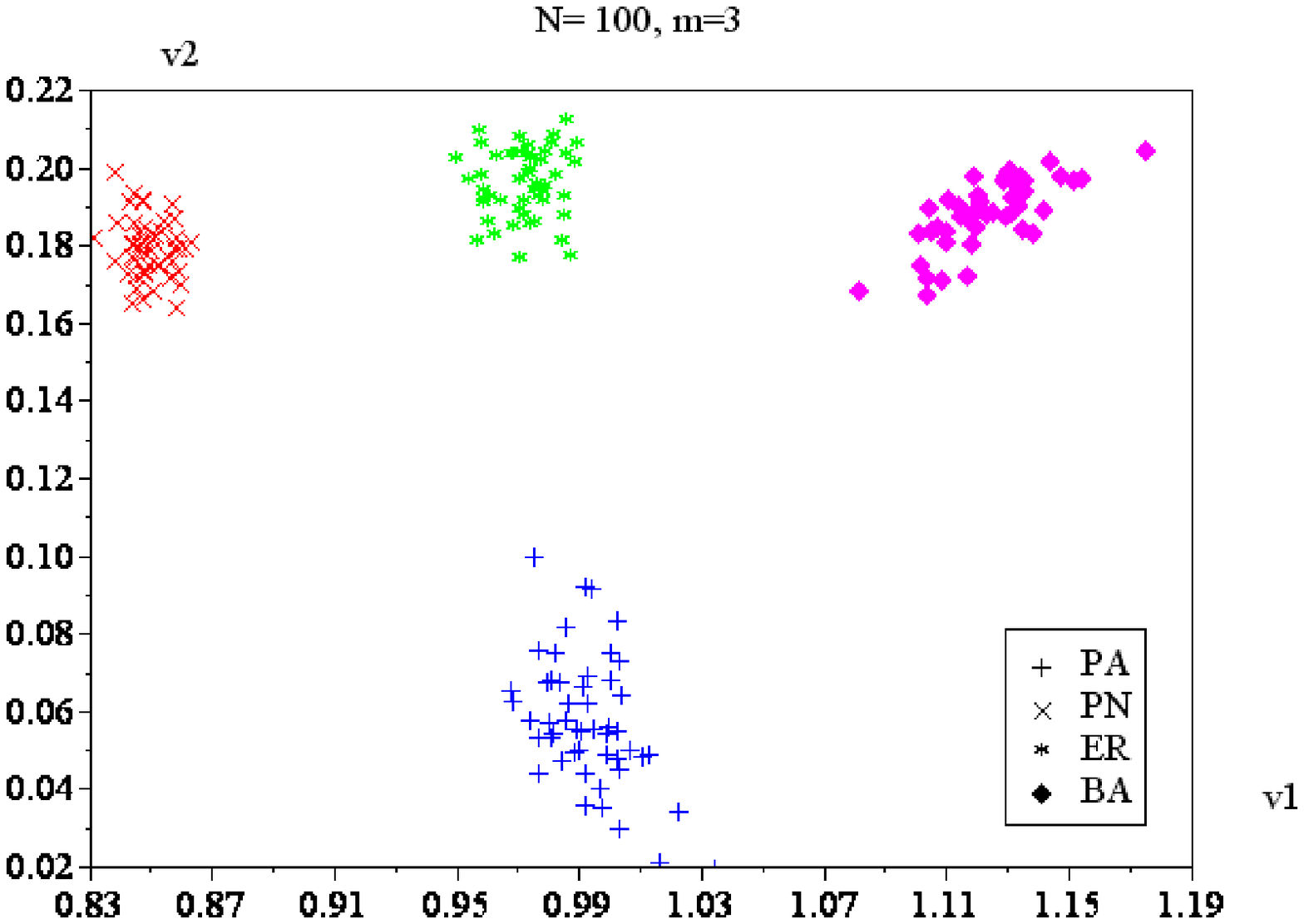}  \hspace{0.5cm}
  \includegraphics[width=0.4\linewidth]{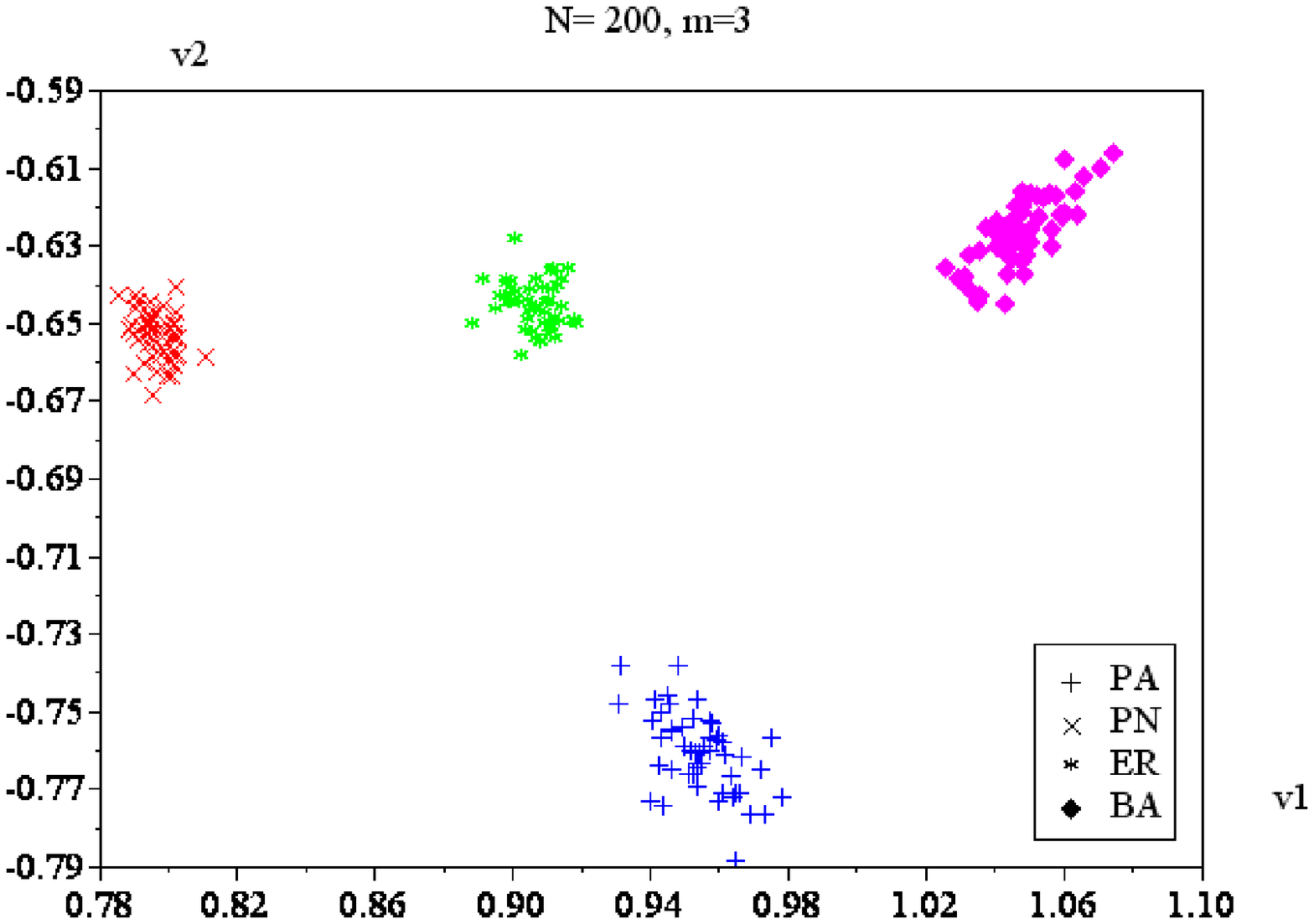}  \\
   (a)  \hspace{7cm}  (b) \\
  \includegraphics[width=0.4\linewidth]{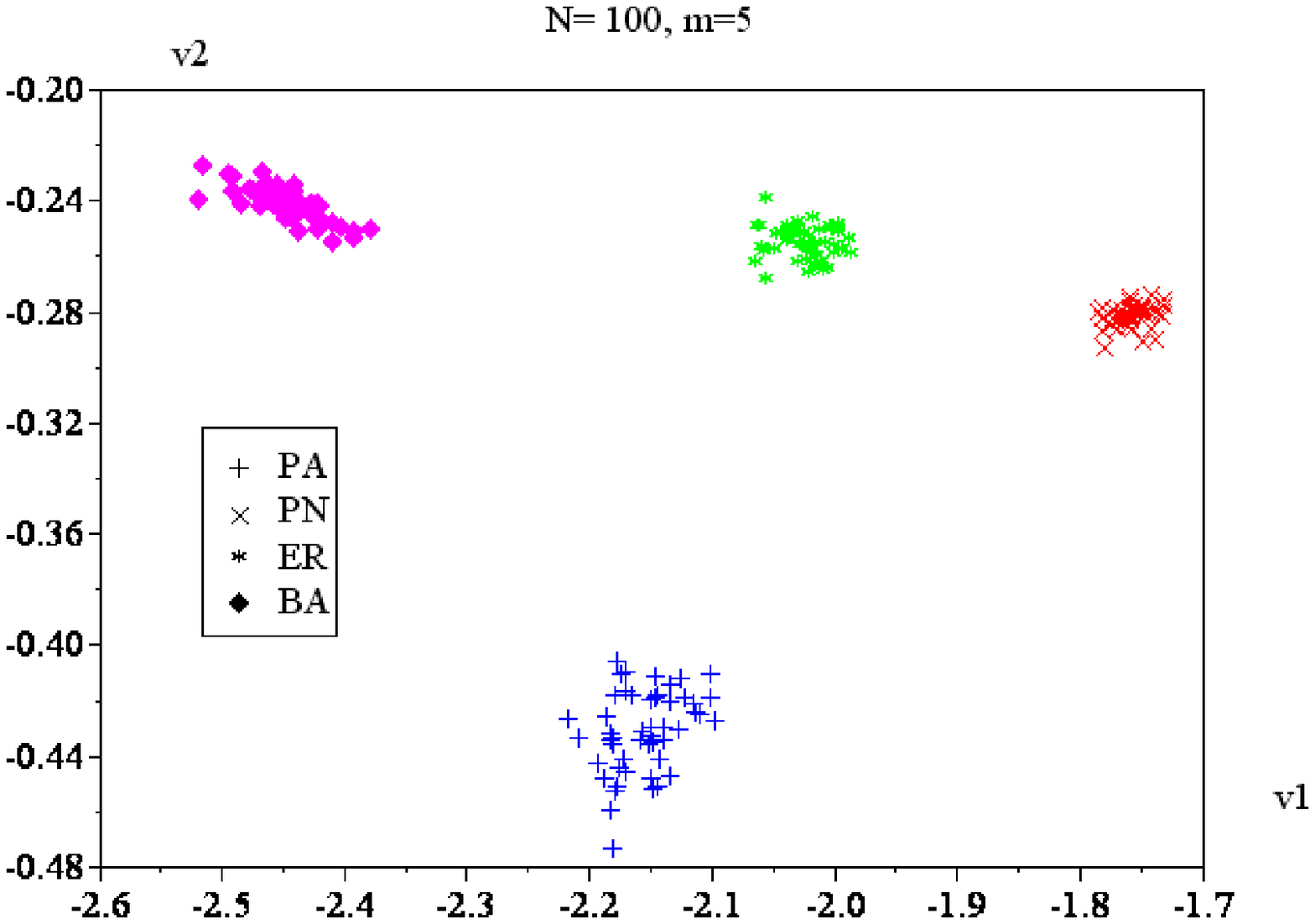}  \hspace{0.5cm}
  \includegraphics[width=0.4\linewidth]{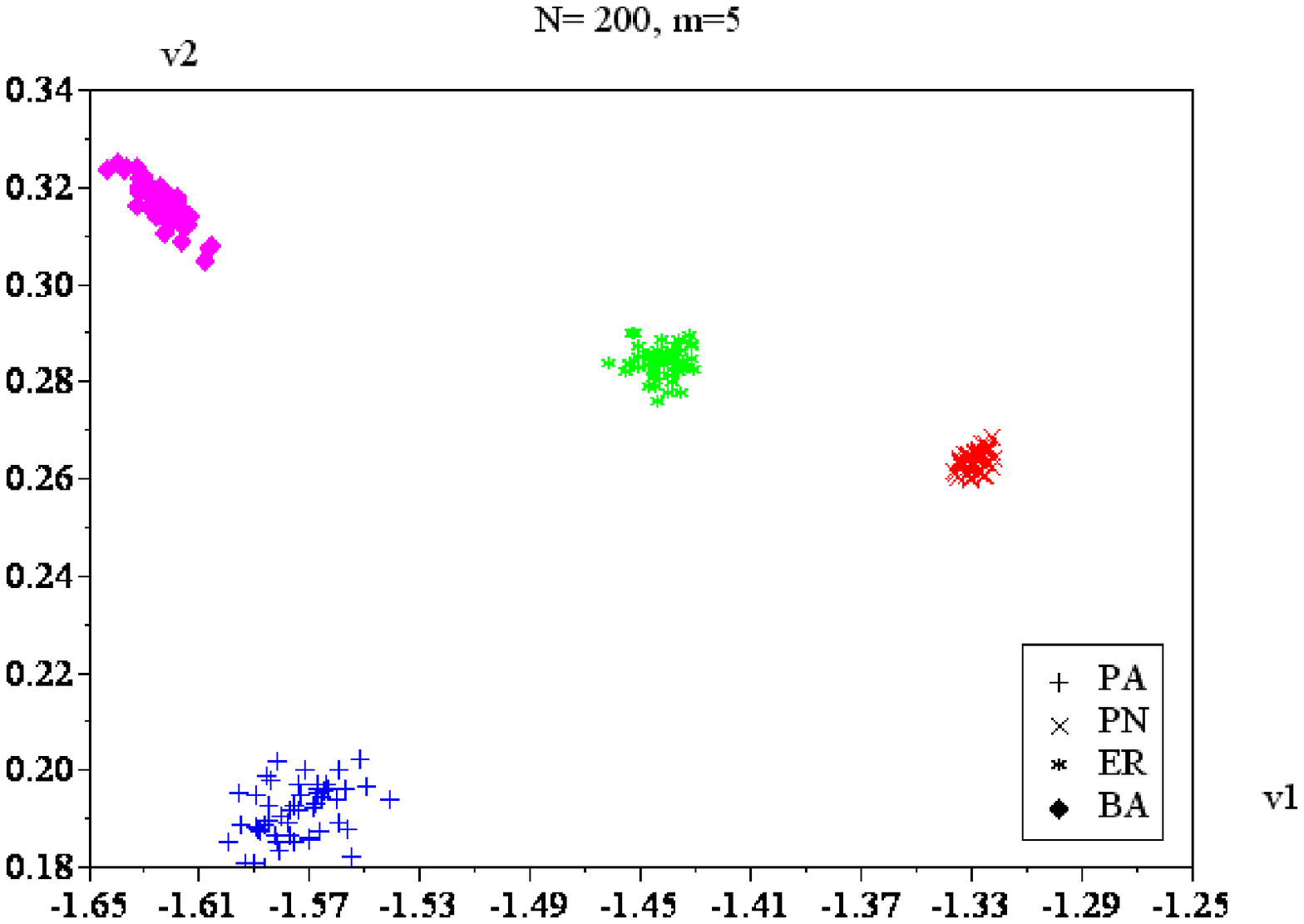}  \\
   (c)  \hspace{7cm}  (d) \\
  \caption{The measurement spaces involving all the measurements
  in Table~\ref{tab:meas} for the
  50 instances of each of the 5 network models considered in this
  article with respect to the four configurations: (a) $N=100$ and
  $m=3$; (b) $N=100$ and $m=5$;(c) $N=200$ and $m=3$; and (d) $N=200$
  and $m=5$. See text for discussion.}~\label{fig:all}
  \end{center}
\end{figure*}

Figure~\ref{fig:all} shows the scatterplots obtained while considering
all the network measurements in Table ~\ref{tab:meas}.  The axes in
these plots refer to the two main canonical variables $v1$ and $v2$,
which are linear combinations of all the considered measurements,
weighted so as to lead to maximum dispersion between the classes and
minimum dispersion inside each of them.  The reader should notice
that, because these axes are obtained from eigenvectors, their
orientation is arbitrary and may reverse from one case to another.  It
is clear from Figure~\ref{fig:all} that the PN, PA, ER and BA
categories of networks yielded well-defined, neat respectie clusters
with relative small inter-cluster dispersion, combined with a strong
disperions among the different classes.  In other words, these results
corroborate the fact that these four models have connectivity
structures which are markedly different one another.  The PN tends to
present the smallest overall dispersion of measurements (specially for
large values of $N$ and $m$), while the PA accounts for the largest
dispersion.

\section{Concluding Remarks}

Two of the most important connectivity patterns (or motifs) underlying
complex networks are paths and stars.  While much of the attention
from the complex networks community has been focused in star
organizations (especially hubs), investigations of the distribution of
paths in networks have been particularly incipient (except mainly for
studies related to betweeness centrality).  The present work has
extended a recent previous investigation~\cite{Costa_path:2007}
involving the duality/transformation between paths and stars in order
to propose a new supercategory of complex networks, namely the
so-called \emph{knitted nets}, corresponding to networks organized in
terms of paths.  Two classes of such networks have been proposed and
investigated: (i) PA, derived from BA networks through the star-path
transformation; and (ii) PN, obtained by randomly selecting all the
network nodes without repetition.  The two main contributions of the
present work include:

{\bf Proposal of a new superclass of complex networks (knitted):}
Because paths can be understood as the dual concept of the star motif
which underlies important models such as scale free networks
(organized in terms of hubs), it is important to consider complex
networks underlain by paths, which have been called \emph{knitted
networks} in the present work.  Each of the two novel types of
networks proposed here present distinctive features.  The PN model,
obtained by performing progressive path-walks involving all nodes
(without repetition), has been found to present remarkable uniformity
of measurements at the individual node level.  This type of network
therefore corresponds to an interesting theoretical model integrating
stochasticity and uniformity.  The second type of proposed networks,
namely the PA nets obtained by path-transformations of BA
counterparts, is potentially interesting because they are formed by a
scale free distribution of path lengths (recall that, in principle,
the path-transformation produces weighted networks).  In this
respect, PA models can be understood as duals of the BA networks.

{\bf Comprehensive characterization of new network models:} Though
several investigations of the structure of complex networks have
considered just a few measurements such as the average node degree,
clustering coefficient and shortest path length, such features are
often unable to comprehensively characterize the connectivity of the
different network types~\cite{Costa_surv:2007}.  This again became
clear in Section~\ref{sec:deg_cc}, where the consideration of only the
average degree and average clustering coefficient did not lead to
significative differences between the PN and PA models and the ER and
BA types of networks.  However, by considering additional
measurements, with emphasis on the standard deviations of traditional
measurements as well as the second clustering coefficient (a
hierarchical or concentric~\cite{Costa:2004, Costa_EPJB:2005,
Costa_JSP:2006, Costa_NJP:2007} kind of measurement), the two new
models PN and PA were shown to be considerably distinct not only one
from the nother, but also from all the four traditional theoretical
models of complex networks presently considered.  The necessity of
dimensionality reduction, accomplished through the multivariated
statistical method of canonical projections, also again confirmed
itself essential for making visual sense of distributions on complets
networks in high dimentionsl spaces.

The concepts and results described in the present work have paved the
way to a number of related future investigations, which include but
are not limited to:

{\bf Maximum path investigations:} By emphasizing the importance of
paths as basic motifs in complex networks, the present work motivates
additional investigations not only on networks underlain by such
motifs, but also in paying greater attention to the path-structure in
all complex networks models.  One point of particular importance which
has been relatively overlooked concerns the statistics of the longest
path between nodes in complex networks.  As a matter of fact, while
great attention has been placed on the analysis of shortes paths, the
study of maximal paths has received scant attention.  Though the
identfication of maximul paths is known to be an NP-complete problem,
the statistics of their lengths in complex networks is poised to
provide valuable insights about diverse theoretical and real networks.

{\bf Considerations of Hierarchical measurements:} A series of
hierarchical/concentric measurements~\cite{Costa:2004} have been
proposed in addition to the second clustering coefficient.  It would
be interesting to consider how the proposed models PN and PA differ
among themselves and with respect to other theretical models as far as
such hierarchical measurements are concerned.

{\bf Additional types of knitted networs:} Though two classes of
knitted networks have been suggested in this work, there are many
other possible structures organized in terms of paths and cycles.  It
would be also interesting to consider networks formed by walks,
instead of paths, allowing nodes to be visited more than once.

{\bf Consideration of directed networks:} Because a great deal of the
attention in complex network research has been placed on undirected
networks, the extension of such approaches --- including that
described in the present article --- to directed complex networks
represent a particularly promising area for future investigations.
Indeed, the PN growing method can be immediately modified in order to
provide directed networks (the edges would be oriented along the order
of node visitation during the progressive path-walks). 

{\bf Transformation of other network models:} Though the current work
concentrated on star-path transformations, it would be interesting to
consider several other possibilities of transformations between
networks models involving motif-transformations.  For instance, it
would be particularly interesting to consider the path-star
transformations described in~\cite{Costa_path:2007} applied to all
paths in the original network.

{\bf Comparison with real-world models:} Though the current work
concentrated in comparisons between the PN/PA structures with
traditional theoretical models, it would be of particular interest to
consider such new networks as putative models of real-world
structures.  After all, one of the main motivations for complex
networks research has been their potential for modeling real-world
structures.  Going back to the scatterplots in Figure~\ref{fig:all},
it would also be very interesting to conceive other network models
capable of filling in the gaps left between the existing models in
those spaces.

\begin{acknowledgments}
Luciano da F. Costa thanks CNPq (308231/03-1) and FAPESP (05/00587-5)
for sponsorship.
\end{acknowledgments}

\bibliography{path_comp}
\end{document}